\documentclass[prb,aps]{revtex4}
\usepackage{graphicx}

\begin{document}

\title{Injection and detection of spin in a semiconductor by tunneling via interface states}
\vspace*{-3mm}
\author{R. Jansen$^1$, A.M. Deac$^2$, H. Saito$^1$ and S. Yuasa$^1$}
\affiliation{$^1\,$National Institute of Advanced Industrial Science and Technology (AIST),
Spintronics Research Center, Tsukuba, Ibaraki, 305-8568, Japan.\\
$^2\,$Helmholtz-Zentrum Dresden-Rossendorf, Institute of Ion Beam Physics and Materials Research,
01314 Dresden, Germany.}

\begin{abstract}
Injection and detection of spin accumulation in a semiconductor having localized states at the
interface is evaluated. Spin transport from a ferromagnetic contact by sequential, two-step
tunneling via interface states is treated not in itself, but in parallel with direct tunneling. The
spin accumulation $\Delta\mu^{ch}$ induced in the semiconductor channel is not suppressed, as
previously argued, but genuinely enhanced by the additional spin current via interface states. Spin
detection with a ferromagnetic contact yields a weighted average of $\Delta\mu^{ch}$ and the spin
accumulation $\Delta\mu^{ls}$ in the localized states. In the regime where
$\Delta\mu^{ls}/\Delta\mu^{ch}$ is largest, the detected spin signal is insensitive to
$\Delta\mu^{ls}$ and the ferromagnet probes the spin accumulation in the semiconductor channel.
\end{abstract}

\maketitle

\indent Spin polarization can be created in non-magnetic
semiconductors by spin-polarized tunneling from a ferromagnetic
contact. This powerful, robust and technologically viable approach
has been demonstrated in various semiconductors, including silicon
and germanium and at room temperature
\cite{tsymbalhandbook,jonker,dash,suzuki,jeon,saitoge,hamayaefield,wangge,jeonge,jain,ibage}.
Considerable discussion has arisen because the magnitude of the
spin accumulation induced in the semiconductor is consistently in
disagreement with the theory for spin injection and spin diffusion
\cite{fertprb,maekawa,fertieee,dery}. The detected spin signal is
often found to be orders of magnitude larger than expected,
particularly for three-terminal devices in which the spin
accumulation is induced and probed by a single magnetic tunnel
contact
\cite{dash,jeon,saitoge,jeonge,jain,ibage,tran,hamayaschottky}.
But in some Si and Ge based devices with non-local geometry (with
separate spin injection and detection contacts) the spin signal is
significantly smaller than predicted if reasonable values of the
contact tunnel spin polarization are used \cite{suzuki,wangge}.
Understanding the origin of these puzzling results is
indispensable because spin injection and detection by a magnetic
tunnel contact is a cornerstone of semiconductor spintronics.\\
\indent While there are indications that the standard theory for
spin injection does not capture all the physics
\cite{dash,saitoge,ibage}, and lateral inhomogeneity of the tunnel
current may also contribute \cite{dash,hamayaschottky}, it is also
heavily debated whether localized states near the semiconductor
interface play a role. These can give rise to resonant tunneling,
non-resonant scattering and inelastic tunneling and thereby reduce
or even invert the tunneling spin polarization
\cite{bratkovsky,zhang,jansendoped1,shang,tsymbal,jansendoped2,jansenresonant,vedyayev,velev,chantis,lu}.
In a different vein, ferromagnet/insulator/semiconductor
structures under photo-excitation were described by sequential,
two-step transport via interface states with their own spin
accumulation and spin relaxation rate
\cite{prinsspin,jansenphoto}. The states are separated from the
ferromagnet by a tunnel barrier and from the semiconductor bulk by
a Schottky barrier and for the latter, transport by thermionic
emission was considered. Just as for spin injection into
non-degenerate semiconductors \cite{jansenprl}, this severely
compromises the spin selectivity of the contacts. Recently, Tran
et al. also considered spin injection by two-step, sequential
transport, but assumed tunneling across the barrier between
localized states and semiconductor \cite{tran}. Importantly, it
was predicted that the spin accumulation $\Delta\mu^{ls}$ in the
localized states can be much larger than the spin accumulation
$\Delta\mu^{ch}$ induced in the
semiconductor channel, albeit under certain conditions.\\
\indent If two-step tunneling via interface states indeed plays a role, it may have crucial
implications for the injection and detection of spin in a multitude of devices that employ tunnel
contacts. Two pertinent questions are: (i) what is the effect of two-step tunneling via interface
states on the spin accumulation that is created in the semiconductor? (ii) how does two-step
tunneling affect the detection of spin accumulation in the semiconductor by a magnetic contact?
Tran et al. predict that the spin accumulation in the semiconductor can be severely suppressed if
spins relax in the intermediate localized states \cite{tran}. They also predict that a
ferromagnetic contact does not probe $\Delta\mu^{ch}$, but instead $\Delta\mu^{ls}$, which can be
much larger than $\Delta\mu^{ch}$, particularly for small density of localized states. Given the
implications, it is unfortunate that it has become practice to automatically attribute enhanced
spin signals seen in experiment to spin accumulation in interface states, without examining whether
the conditions to produce an enhancement are fulfilled, and without critical tests, for instance,
varying specific parameters and observing whether the experimental data follows the expected trends.\\
\indent To address the effect of interface states, a correct
prediction of their impact on spin transport is required. It is
shown here that Tran's model \cite{tran} and the trends it
predicts need significant revision, because a basic assumption,
namely that {\em all} the tunnel current between ferromagnet and
semiconductor is through localized states, is not generally valid.
Here we treat two-step tunneling via interface states {\em in
parallel with direct tunneling}. We show that the spin
accumulation in the semiconductor channel is not suppressed, but
genuinely enhanced by the additional spin current via interface
states. We also find that spin detection with a ferromagnetic
contact yields a weighted average of $\Delta\mu^{ch}$ and
$\Delta\mu^{ls}$, which shifts depending on the ratio of direct
and two-step tunneling current. Spin accumulation in interface
states only enhances the detected spin signal in the intermediate
regime where both current components are comparable, and only if
the localized states are separated from the semiconductor by a
barrier with sufficiently large resistance.

\begin{figure}[htb]
\hspace*{0mm}\includegraphics*[width=75mm]{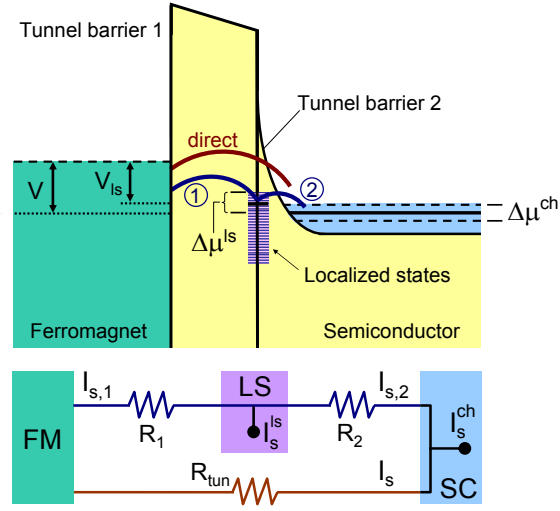}
\caption{Energy band diagram of a
ferromagnet/insulator/semiconductor junction with localized states
(LS) separated from the ferromagnet (FM) by tunnel barrier 1, and
from the semiconductor (SC) by tunnel barrier 2. A spin
accumulation exists in the semiconductor channel
($\Delta\mu^{ch}$) and in the localized states ($\Delta\mu^{ls}$).
The circuit depicts the spin currents and resistances, with
$\bullet$ representing spin sinks due to spin relaxation in LS and
SC. The FM is the spin source.} \label{fig1}
\end{figure}

\indent The system has three sections, a ferromagnet, localized
states with spin-integrated density $D_{ls}$ (in states
eV$^{-1}$m$^{-2}$) and a semiconductor channel (Fig. 1). The
potential of the ferromagnet is taken to be the zero. The
spin-averaged potentials of the semiconductor channel and the
localized states are $V$ and $V^{ls}$, respectively. For direct
tunneling, the charge current $I$ and spin current $I_s$ are (see
also appendix):
\begin{eqnarray}
I = G\,V - P_G\,G\,\left(\frac{\Delta\mu^{ch}}{2}\right) \label{eq1} \\
I_s = P_G\,G\,V - G \left(\frac{\Delta\mu^{ch}}{2}\right) \label{eq2}
\end{eqnarray}
where $G$ is the total (spin-averaged) tunnel conductance and
$P_G$ is the spin polarization of the conductance. Note that the
spin accumulation decays into the semiconductor and that
$\Delta\mu^{ch}$ is the value at the interface, since this
determines the tunneling process. For two-step tunneling via
localized states, we denote the charge and spin current between
ferromagnet and localized states by $I_{1}$ and $I_{s,1}$,
respectively, the total conductance by $G_{1}$ and the conductance
spin polarization by $P_{G1}$. For the second tunnel step between
localized states and semiconductor channel, the charge and spin
current are denoted by $I_{2}$ and $I_{s,2}$, the total tunnel
conductance is $G_{2}$, and the conductance is unpolarized since
neither localized states nor semiconductor is ferromagnetic. The
charge and spin currents for two-step tunneling are (see also the
appendix):
\begin{eqnarray}
I_{1} =  G_{1}\,V^{ls} - P_{G1}\,G_{1}\left(\frac{\Delta\mu^{ls}}{2}\right) \label{eq3} \\
I_{s,1} = P_{G1}\,G_{1}\,V^{ls} - G_{1}\,\left(\frac{\Delta\mu^{ls}}{2}\right) \label{eq4} \\
I_{2} = G_{2}\,(V-V^{ls}) \label{eq5} \\
I_{s,2} = G_{2}\left(\frac{\Delta\mu^{ls}-\Delta\mu^{ch}}{2}\right) \label{eq6}
\end{eqnarray}
Since direct and two-step tunneling occur in parallel, $\Delta\mu^{ch}$ is determined by the total
spin current $I_s+I_{s,2}$ into the channel, where $I_{s,2}$ is proportional to the difference
between $\Delta\mu^{ls}$ and $\Delta\mu^{ch}$. The spin accumulation in the localized states gives
rise to spin relaxation and an associated spin current
$I_{s}^{ls}=e(N_{ls}^{\uparrow}-N_{ls}^{\uparrow})/\tau_s^{ls}$, where $N_{ls}^{\sigma}$ is the
number of electrons with spin $\sigma$ in the localized states, and $\tau_s^{ls}$ is the
spin-relaxation time in the localized states. Note that $I_{s}^{ls}$ is defined in units of
electron angular momentum $\hbar/2$ transferred per unit time, instead of spin flips per unit time.
The spin resistance of the localized states is $r_s^{ls}=\tau_s^{ls}/(e\,D_{ls})$, such that
$\Delta\mu^{ls} = 2\,I_{s}^{ls}\,r_{s}^{ls}$. Similarly, spin relaxation in the semiconductor
channel produces a spin-relaxation spin current $I_{s}^{ch}$ that is related to the spin
accumulation by the spin resistance $r_s^{ch}$ of the semiconductor: $\Delta\mu^{ch} =
2\,I_{s}^{ch}\,r_{s}^{ch}$. The relations for $r_s^{ls}$ and $r_s^{ch}$, together with
eqns.(\ref{eq1})-(\ref{eq6}), define the system. The three unknown quantities ($\Delta\mu^{ch}$,
$\Delta\mu^{ls}$ and $V^{ls}$) are obtained from the following three conditions: (i) $I_{s}^{ch} =
I_{s} + I_{s,2}$, (ii) $I_{s}^{ls} = I_{s,1} - I_{s,2}$, and (iii) $I_{1} = I_{2}$. Condition (i)
says that in a steady state, the spin relaxation spin current in the semiconductor is equal to the
total spin current injected into it (sum of $I_{s}$ and $I_{s,2}$). Condition (ii) states that the
spin relaxation spin current in the localized states must be equal to the difference of the spin
current $I_{s,1}$ injected into it from the ferromagnet and the spin current $I_{s,2}$ that leaks
away into the semiconductor. Charge conservation yields condition (iii). The solutions for the spin
accumulations are \cite{note1}:
\begin{equation}
\Delta\mu^{ls}\, =
\frac{\beta^{ch}P_{G1}+P_{G}}{\beta^{ch}\beta^{ls}-1}\left(\frac{2\,R_2}{R_{tun}}\right)\,V
\,\,\,\,\,\,\,\,\, \label{eq7}
\end{equation}
\begin{equation}
\Delta\mu^{ch} =
\frac{\beta^{ls}P_{G}\,+P_{G1}}{\beta^{ch}\beta^{ls}-1}\left(\frac{2\,R_2}{R_1+R_2}\right)\,V
\label{eq8}
\end{equation}
where we defined the resistances $R_{tun}=1/G$, $R_{1}=1/G_1$, $R_{2}=1/G_2$ and the dimensionless
parameters:
\begin{equation}
\beta^{ch} = \frac{R_{tun}\,R_2+r_s^{ch}\,(R_2+R_{tun})}{r_s^{ch}\,(R_1+R_2)} \approx
\frac{R_{tun}\,(R_2+r_s^{ch})}{R_1\,r_s^{ch}} \label{eq9}
\end{equation}
\begin{equation}
\beta^{ls} = \frac{R_1\,R_2\,(R_1+R_2)+r_s^{ls}\,[(R_1
+R_2)^2-(P_{G1}\,R_2)^2]}{r_s^{ls}\,R_1\,R_{tun}} \approx
\frac{R_{1}\,(R_2+r_s^{ls})}{R_{tun}\,r_s^{ls}} \label{eq10}
\end{equation}
The approximate forms of $\beta^{ch}$ and $\beta^{ls}$ are obtained when $R_1>>R_2$, which applies
to localized states at or near the semiconductor interface. If $R_1>>R_2$, eqns. (\ref{eq7}) and
(\ref{eq8}) reduce to:
\begin{equation}
\Delta\mu^{ls} =
 \left(\frac{2\,r_s^{eff}}{R_1}\right)\,P_{G1}\,V +
\frac{r_s^{eff}}{R_2+r_s^{ch}}\left(\frac{2\,r_s^{ch}}{R_{tun}}\right)\,P_{G}\,V \label{eq11}
\end{equation}
\begin{equation}
\Delta\mu^{ch} = \frac{r_s^{ch}}{R_2+r_s^{ch}}\left(\frac{2\,r_s^{eff}}{R_1}\right)\,P_{G1}\,V +
\frac{r_s^{ls}+R_2}{r_s^{ls}+R_2+r_s^{ch}}\left(\frac{2\,r_s^{ch}}{R_{tun}}\right)\,P_{G}\,V
\label{eq12}
\end{equation}
where $r_s^{eff} = r_s^{ls}\,(R_{2} + r_s^{ch})/(r_s^{ls} + R_{2} + r_s^{ch})$ as in the work of
Tran et al. \cite{tran}. It represents the effective spin resistance of the system of localized
states and semiconductor channel, coupled by a tunnel resistance $R_{2}$.\\
\indent The spin accumulations have a contribution from two-step tunneling (proportional to
$P_{G1}$) and a contribution that arises from direct tunneling (proportional to $P_{G}$). The
latter disappears for $R_{tun}\rightarrow\infty$, for which eqns. (\ref{eq11}) and (\ref{eq12})
reduce to that obtained in Tran's model \cite{tran}. In that case one finds that the spin
accumulation is governed by $r_s^{eff}$ instead of $r_s^{ch}$, and that
$\Delta\mu^{ls}/\Delta\mu^{ch}$ equals $1+R_2/r_s^{ch}$, which can be much larger than unity when
$R_2>r_s^{ch}$. Moreover, $\Delta\mu^{ch}$ becomes vanishingly small when $R_2>r_s^{ls},r_s^{ch}$,
corresponding to the situation where spins relax in the localized states before escaping into the
semiconductor. In Tran's model, a spin current into the semiconductor is obtained only when spin
relaxation in localized states is negligible ($R_2<r_s^{ls}$).\\
\indent The behavior changes drastically when direct tunneling is included (finite $R_{tun}$). The
spin current injected into the semiconductor by direct tunneling is approximately $P_G\,V/R_{tun}$,
and the associated contribution to $\Delta\mu^{ch}$ (last term in eqn. (\ref{eq12})) exists in
addition to the two-step tunneling contribution. In other words, starting with direct tunneling at
a given bias voltage $V$ and then adding localized states, one increases $\Delta\mu^{ch}$, since
extra spin current is injected into the semiconductor by the two-step tunneling. This extra current
can also be highly spin polarized (for $R_2<r_s^{ls}$), which is beneficial for creating a large
spin accumulation in the semiconductor channel. Even if the spin current from the localized states
is negligible (when $R_2>r_s^{ls},r_s^{ch}$), the spin accumulation induced by direct tunneling
still remains. Our formalism thus demonstrates that neglecting direct tunneling leads to an
incorrect prediction of the magnitude of $\Delta\mu^{ch}$ and to the erroneous conclusion that
localized states have a detrimental effect on the spin accumulation in the semiconductor channel.
Treating direct and two-step tunneling on an equal
footing is thus crucial in order to assess how localized states affect the induced spin polarization.\\
\indent Next we address how two-step tunneling via interface states affects the {\em detection} of
a spin accumulation in the semiconductor. Spin detection is typically done by suppressing the spin
accumulation via spin precession in a magnetic field perpendicular to the injected spins (Hanle
effect). At constant charge current, the resulting change in voltage $\Delta V_{Hanle}$ across the
tunnel contact is, without approximations:
%\begin{equation}
%\Delta V_{Hanle} = \left(\frac{R_2}{R_1+R_2+R_{tun}}\right)
%\left\{\frac{\beta^{ls}(P_{G})^2+\beta^{ch}(P_{G,1})^2+2\,P_G\,P_{G,1}}{\beta^{ch}\beta^{ls}-1}\right\}\,V
%\label{eq69c}
%\end{equation}
\begin{equation}
\Delta V_{Hanle} = \frac{R_1+R_2}{R_1+R_2+R_{tun}}\left(\frac{P_G}{2}\right)\Delta\mu^{ch} +
\frac{R_{tun}}{R_1+R_2+R_{tun}}\left(\frac{P_{G1}}{2}\right)\Delta\mu^{ls} \label{eq13}
\end{equation}
where $\Delta\mu^{ch}$ and $\Delta\mu^{ls}$ are the values in the
absence of a magnetic field (eqns.(\ref{eq7}) and (\ref{eq8})).
The important point is that the Hanle signal is a {\em weighted
average} of $\Delta\mu^{ch}$ and $\Delta\mu^{ls}$, with a relative
contribution determined by the ratio of the resistances associated
with direct tunneling ($R_{tun}$) and two-step tunneling
($R_1+R_2$). When the current is dominated by the localized states
($R_{tun}>>R_1+R_2$), the first term is zero and the Hanle signal
is governed exclusively by $\Delta\mu^{ls}$, as in Tran's model
\cite{tran}. However, when the current due to two-step tunneling
is comparable to or smaller than the direct tunneling current, the
weight shifts to the term proportional to $\Delta\mu^{ch}$ and any
enhancement of the Hanle signal due to localized states
disappears. The resistance of the junction is then determined by
direct tunneling, and $\Delta V_{Hanle}$ is insensitive to
$\Delta\mu^{ls}$ (a large $\Delta\mu^{ls}$ may still exist, but
the voltage across the junction does not depend on it). This
essential behavior is not captured when one considers only
two-step tunneling.

\begin{figure}[htb]
\hspace*{0mm}\includegraphics*[width=75mm]{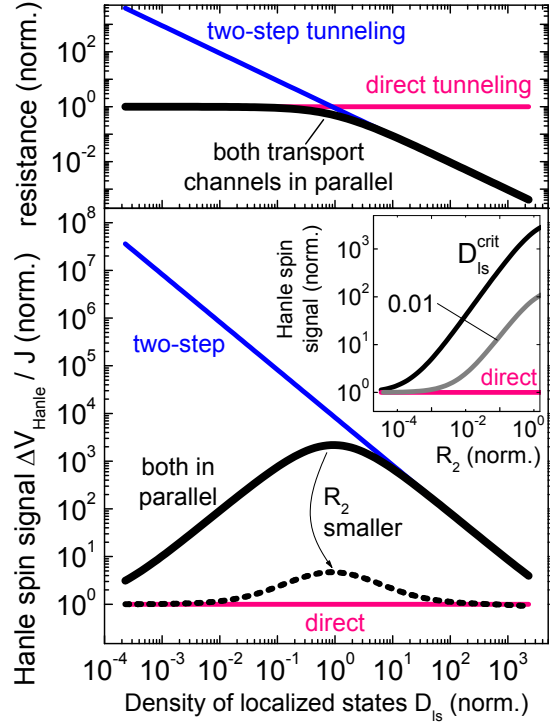}
\caption{Tunnel resistance (top) and spin signal $\Delta
V_{Hanle}$ divided by the current density $J$ (bottom) as a
function of the density of localized states $D_{ls}$, for pure
two-step tunneling (blue), pure direct tunneling (pink), and for
two-step tunneling and direct tunneling in parallel (black - the
dotted line is for $\tau_2^{esc}$ and thus $R_2$ reduced by a
factor of 1000). The horizontal axis is normalized to the value of
$D_{ls}$ for which the currents by direct and two-step tunneling
are equal. The top and bottom vertical axes are normalized to,
respectively, $R_{tun}$ and the spin signal for pure direct
tunneling. The escape times $\tau_1^{esc}$, $\tau_2^{esc}$ as well
as $R_{tun}$ were taken to be independent of $D_{ls}$. The inset
displays the spin signal versus $R_2$ for $D_{ls}=D_{ls}^{crit}$
and $D_{ls}=0.01\,D_{ls}^{crit}$.} \label{fig2}
\end{figure}

\indent For a given tunnel barrier, the relative weight of direct
and two-step tunneling is proportional to the density of localized
states because $R_1$ and $R_2$ scale inversely with $D_{ls}$. This
can be seen by writing $R_{1} = \tau_1^{esc}/(e\,D_{ls})$ and
$R_{2} = \tau_2^{esc}/(e\,D_{ls})$, where $\tau_1^{esc}$ and
$\tau_2^{esc}$ are the characteristic time for escape of an
electron from, respectively, localized states into the ferromagnet
and into the semiconductor channel, as determined by the
transmission probability of tunnel barrier 1 and 2. At large
$D_{ls}$, the resistance for two-step tunneling is smaller than
the resistance for direct tunneling (top panel of Fig.
\ref{fig2}). As $D_{ls}$ is reduced, $R_1+R_2$ increases rapidly
and surpasses $R_{tun}$ at a critical value $D_{ls}^{crit}$.
Beyond this, direct tunneling dominates. This has a marked effect
on the Hanle signal (bottom panel). Tran's model \cite{tran}
predicts increasingly large values of the Hanle signal at smaller
$D_{ls}$ (blue curve) because lower $D_{ls}$ means larger spin
resistance ($r_s^{ls}\propto 1/D_{ls}$) and thus a larger spin
accumulation in the localized states. However, our full model
shows that the Hanle signal goes through a maximum at
$D_{ls}^{crit}$, and for smaller $D_{ls}$, the Hanle signal is
reduced and approaches the value obtained for pure direct
tunneling. We thus find that Tran's model does not predict the
correct variation with $D_{ls}$ and is not valid in the regime
where it predicts the largest enhancement of the spin signals - it
grossly overestimates the Hanle signal for $D_{ls}<D_{ls}^{crit}$.
For large $D_{ls}$, where the signal enhancement is limited,
Tran's model gives approximately the correct value of the Hanle
signal, but note that even then it does not predict the correct
value of $\Delta\mu^{ch}$, as explained. The value of
$D_{ls}^{crit}$ depends on the tunnel probabilities for direct and
two-step tunneling through the condition $R_{tun}\approx
(\tau_1^{esc}+\tau_2^{esc})/(e\,D_{ls}^{crit})$.\\
\indent Finally, we discuss an important and often overlooked
characteristic of two-step tunneling. The value of $r_s^{eff}$
(which governs $\Delta\mu^{ls}$) can be much larger than
$r_s^{ch}$, but $r_s^{eff}$ cannot be larger than $R_2$. A smaller
$R_2$ means a stronger coupling between localized states and
semiconductor channel, which tends to equalize their spin
accumulations and suppress $\Delta\mu^{ls}$. Hence, any
enhancement of the Hanle signal by localized interface states, if
present, can be suppressed by reducing $R_2$, i.e., by reducing
the energy barrier that separates localized states from the
semiconductor bulk. For example, when $\tau_2^{esc}$ and thus
$R_2$ is reduced by a factor of 1000 at fixed $D_{ls}$, the
maximum Hanle signal is also reduced by about the same factor
(Fig. 2, dotted black curve, and inset). Moreover, enhancement
becomes limited to a narrower interval around $D_{ls}^{crit}$.
This feature was exploited in the experiments by Dash et al. to
exclude interface states as a source of the large spin
accumulation observed in silicon at room temperature \cite{dash}.
They used a treatment with Cs to reduce the Schottky barrier, but
found spin signals to remain large and much larger than can be
supported by the small Schottky barrier (small $R_2$). We suggest
that if spin signals are observed that exceed the predictions of
spin injection theory, one must look beyond the magnitude of the
signal and investigate trends in order to determine whether an
enhancement due to localized states is at play. The model
presented here describes how two-step tunneling via localized
interface states affects the injection and the detection of spin
with a ferromagnetic contact, and the resulting trends, providing
a firm basis for comparison with experiments.

\begin{appendix}
\section{Currents, potentials and Hanle signal}
In this appendix we provide the equations for the current by direct and two-step
tunneling for each spin orientation separately. For the sake of
completeness, we also provide the full solutions for the potentials and
the Hanle signals, without approximations.\\
\\
\indent For direct tunneling between ferromagnet and semiconductor
channel, we denote the tunnel currents of majority ($\uparrow$)
and minority ($\downarrow$) spin electrons by $I^{\uparrow}$ and
$I^{\downarrow}$, respectively, and the corresponding tunnel
conductances by $G^{\uparrow}$ and $G^{\downarrow}$. With the
voltage definitions described in the main text we have:
\begin{eqnarray}
I^{\uparrow} = G^{\uparrow}\left(V-\frac{\Delta\mu^{ch}}{2}\right) \label{eq22}\\
I^{\downarrow} =
G^{\downarrow}\left(V+\frac{\Delta\mu^{ch}}{2}\right) \label{eq23}
\end{eqnarray}
The charge tunnel current $I= I^{\uparrow}+I^{\downarrow}$ and the
spin tunnel current $I_s= I^{\uparrow}-I^{\downarrow}$ due to
direct tunneling are then:
\begin{eqnarray}
I = G\,V - P_G\,G\,\left(\frac{\Delta\mu^{ch}}{2}\right) \label{eq24} \\
I_s = P_G\,G\,V - G \left(\frac{\Delta\mu^{ch}}{2}\right)
\label{eq25}
\end{eqnarray}
with the total conductance $G = G^{\uparrow}+G^{\downarrow}$ and
the tunnel spin
polarization $P_G = (G^{\uparrow}-G^{\downarrow})/(G^{\uparrow}+G^{\downarrow})$.\\
\indent For two-step tunneling via localized interface states, we
denote the tunnel currents between ferromagnet and localized
states of majority and minority spin electrons by
$I_{1}^{\uparrow}$ and $I_{1}^{\downarrow}$, respectively, and the
corresponding tunnel conductances by $G_{1}^{\uparrow}$ and
$G_{1}^{\downarrow}$. Tunneling between localized states and
semiconductor channel is described by the tunnel currents
$I_{2}^{\uparrow}$ and $I_{2}^{\downarrow}$, and a tunnel
conductance $G_{2}/2$ per spin. The latter is independent of spin
because the semiconductor and the localized states are both not
ferromagnetic. The tunnel current components for two-step
tunneling via localized states are:
\begin{eqnarray}
I_{1}^{\uparrow} = G_{1}^{\uparrow}\left(V_{ls}-\frac{\Delta\mu^{ls}}{2}\right) \label{eq14}\\
I_{1}^{\downarrow} = G_{1}^{\downarrow}\left(V_{ls}+\frac{\Delta\mu^{ls}}{2}\right) \label{eq15}\\
I_{2}^{\uparrow} =
\frac{G_{2}}{2}\left(V-V_{ls}+\left(\frac{\Delta\mu^{ls}-\Delta\mu^{ch}}{2}\right)\right)  \label{eq16}\\
 I_{2}^{\downarrow} = \frac{G_{2}}{2}\left(V-V_{ls}-\left(\frac{\Delta\mu^{ls}-\Delta\mu^{ch}}{2}\right)\right) \label{eq17}
\end{eqnarray}
The charge tunnel current $I_{1} =
I_{1}^{\uparrow}+I_{1}^{\downarrow}$ and spin tunnel current
$I_{s,1} = I_{1}^{\uparrow}-I_{1}^{\downarrow}$ between
ferromagnet and localized states are then:
\begin{eqnarray}
I_{1} =  G_{1}\,V_{ls} - P_{G1}\,G_{1}\left(\frac{\Delta\mu^{ls}}{2}\right) \label{eq18} \\
I_{s,1} = P_{G1}\,G_{1}\,V_{ls} -
G_{1}\,\left(\frac{\Delta\mu^{ls}}{2}\right) \label{eq19}
\end{eqnarray}
with $G_{1} = G_{1}^{\uparrow}+G_{1}^{\downarrow}$ and $P_{G1} =
(G_{1}^{\uparrow}-G_{1}^{\downarrow})/(G_{1}^{\uparrow}+G_{1}^{\downarrow})$.
The charge tunnel current $I_{2} =
I_{2}^{\uparrow}+I_{2}^{\downarrow}$ and spin current $I_{s,2} =
I_{2}^{\uparrow}-I_{2}^{\downarrow}$ between localized states and
semiconductor are:
\begin{eqnarray}
I_{2} = G_{2}\,(V-V_{ls}) \label{eq20} \\
I_{s,2} =
G_{2}\left(\frac{\Delta\mu^{ls}-\Delta\mu^{ch}}{2}\right)
\label{eq21}
\end{eqnarray}
Equations (\ref{eq24}), (\ref{eq25}) and (\ref{eq18})-(\ref{eq21})
are given in the main text.\\
\\
\indent The solution for the potential of the localized states is:
\begin{equation}
V^{ls} = \left(\frac{R_1}{R_1+R_2}\right)
\left\{1+\left(\frac{(R_2)^2}{R_1\,R_{tun}}\right)\left\{\frac{\beta^{ch}P_{G1}+P_{G}}{\beta^{ch}\beta^{ls}-1}\right\}\right\}\,V
\end{equation}
The voltage across the tunnel contact is related to the total
current $I^{tot}=I+I_{2}$ by:
\begin{equation}
V =
\frac{R_1+R_2}{(R_1+R_2+R_{tun})-R_2\,\left\{\frac{\beta^{ls}(P_{G})^2+\beta^{ch}(P_{G1})^2+2\,P_G\,P_{G1}}{\beta^{ch}\beta^{ls}-1}\right\}}\,R_{tun}\,I^{tot}
\end{equation}
The full expressions for the Hanle signal in terms of $V$ or
$I^{tot}$ are:
\begin{equation}
\Delta V_{Hanle} = \left(\frac{R_2}{R_1+R_2+R_{tun}}\right)
\left\{\frac{\beta^{ls}(P_{G})^2+\beta^{ch}(P_{G1})^2+2\,P_G\,P_{G1}}{\beta^{ch}\beta^{ls}-1}\right\}\,V
\label{eq69c}
\end{equation}
\begin{equation}
\Delta V_{Hanle} =
\left\{\frac{\left(\frac{R_2\,(R_1+R_2)}{R_1+R_2+R_{tun}}\right)}{(R_1+R_2+R_{tun})\,\left\{\frac{\beta^{ch}\beta^{ls}-1}{\beta^{ls}(P_{G})^2+\beta^{ch}(P_{G1})^2+2\,P_G\,P_{G1}}\right\}
- R_2}\right\}\,R_{tun}\,I^{tot}
\end{equation}

\end{appendix}

%\clearpage


\begin{thebibliography}{99}
\bibitem{note1} It was confirmed that eqns. (\ref{eq7}) and (\ref{eq8}) reduce to the results for pure
direct tunneling and pure two-step tunneling in the appropriate limits, respectively,
$R_{tun}<<R_1+R_2$ and $R_{tun}>>R_1+R_2$. Note that setting $P_G$ (or $P_{G1}$) to zero is not
sufficient, as this does not remove the charge current and also not the complete spin current.
\bibitem{note2} The term proportional to $P_G$ in eqn. (\ref{eq11}) implies that $\Delta\mu^{ls}$
is also enhanced by including direct tunneling. This is because it increases $\Delta\mu^{ch}$,
which tends to reduce the spin current $I_{s,2}$ between localized states and the channel since
this is proportional to $\Delta\mu^{ls}-\Delta\mu^{ch}$. The spin accumulation in the localized
states then has to increase to restore the balance of the spin currents going in and out. The
relative effect is small when $\Delta\mu^{ls}>>\Delta\mu^{ch}$.
\end{thebibliography}
\end{document}